# On the strong spherical shock waves in a two-phase gas-particle medium

**R. K. Anand**

**Department of Physics, University of Allahabad, Allahabad-211002, India**

e-mail address: anand.rajkumar@rediffmail.com

**Abstract**

In this paper, power series solutions for strong spherical shocks of time dependent variable energy propagating in a two-phase gas-particle medium are presented taking into consideration the power series solution technique (Sakurai in J Phys Soc Jpn 8:662–669, 1953; Freeman in J Phys D Appl Phys 2(1):1697-1710, 1968). Assuming the medium to be a mixture of a perfect gas and small solid particles, the power series solutions are obtained in terms of $M^{-2}$, where $M$ is the upstream Mach number of shock. This investigation presents an overview of the effects due to an increase in (i) the propagation distance from the inner expanding surface and, (ii) the dust loading parameters on flow-field variables such as the velocity of fluid, the pressure, the density, and also on the speed of sound, the adiabatic compressibility of mixture and the change-in-entropy behind the strong spherical shock front.

## Introduction

In 1942 Guderley [1] first obtained self-similar solutions describing a converging strong shock wave propagating in an ideal gas. Such families of solutions require invariant boundary conditions under the similarity transformation. Van Dyke and Guttmann [2] described a converging shock driven by a piston with the help of analytical series. In series, the zeroth order term corresponds to the plane problem and the higher order terms account for the spherical effects. Oshima [3] described a diverging shock wave with approximations valid in three domains depending on the Mach number: strong, intermediate, and weak shock. Sakurai [4] presented another method to describe the diverging shocks and obtained the solutions in power series of $M^{-2}$, where $M$ is the upstream shock Mach number. Sakurai's power series solutions are for initially strong shock waves of constant energy and the zeroth order term of series solutions corresponds to the self-similar solutions for the Taylor-Sedov problem of a point explosion. The mathematical proof of existence of this solution has been demonstrated in 2009 by Takahashi [5]. Hafner [6] presented a power series solution for strong converging shock waves near the centre of convergence. The series form generalization of Guderley's imploding shock problem was suggested by Hunter [7] and done later by Welsh [8]. In a same way, Ponchaut et al. [9] extended the self-similar Guderley's solution by using series form solutions and later on Hornung et al. [10] obtained a universal solution for converging shock waves.

Gretler and Regenfelder [11] presented a similarity solution for strong blast waves of variable energy propagating in a dusty gas.

In this paper, the problem studied by Gretler and Regenfelder [11] has been revisited and the solutions for spherical shock wave with time dependent variable energy propagating in two-phase gas-particle medium has been presented in the form of power series. The power series solutions have been obtained in terms of $M^{-2}$, where $M$ is the upstream shock Mach number, taking into consideration the power series solution technique [4,12]. On the original idea of Pai [13] we have assumed the medium as a mixture of a perfect gas and a pseudo-fluid of solid particles at a velocity and temperature equilibrium with a constant ratio of specific heat of the mixture. To ensure that the two-phase gas-particle medium is physically reasonable, both kinematic and thermal equilibrium must hold i.e., both Stokes numbers should be less than $10^{-3}$ [14].

This work presents an overview of the effects due to an increase in (i) the propagation distance from the inner expanding surface and, (ii) the dust loading parameters on flow-field variables such as the velocity of fluid, the pressure, the density, and also on the speed of sound, the adiabatic compressibility of mixture and the change-in-entropy behind the strong spherical shock front. The results are displayed graphically and discussed by comparison with the previous investigations for an ideal gas, i.e. dust-free gas as limiting cases. The analysis presented can give results substantially different from the ideal gas solution (with modified thermodynamic constants) only when the volumetric concentration is greater than $10^{-3}$ [15]. The paper is organized as follows: The background information is provided in "Introduction" section. "Equations of Motion and Boundary Conditions" contains general assumptions and notations, basic equations and boundary conditions. In "Power Series Solutions" section the methodology of obtaining power series solutions is described. "Results and Discussion" section mainly presents results with discussion on the important components of the present model. The findings are concluded in "Conclusions" section with details on which effects were accounted for and which were not.

## Equations of Motion and Boundary Conditions

In our investigation the total energy of the flow-field behind the shock front is time dependent and varying according to a power law [11,12,16,17] of the form $E = E_o t^k$, where $E_o$ is a functional constant and $k \geq 0$ is energy-input parameter. It is notable that $k = 0$ corresponds to the instantaneous constant energy blast wave, whereas $k > 0$ corresponds to the case in which the total energy increases with time. The non-dimensional form of the conservation equations governing an unsteady, spherically symmetric flow-field between spherical shock front and inner expanding surface moving in a two-phase gas-particle medium can be expressed as [11]:

$$\lambda y \frac{\partial h}{\partial y} + (f - x)\frac{\partial h}{\partial x} + h\left(\frac{\partial f}{\partial x} + \frac{2f}{x}\right) = 0 \qquad (1a)$$

$$\lambda y\frac{\partial f}{\partial y}+(f-x)\frac{\partial f}{\partial x}-\frac{\lambda}{2}f+\frac{1}{h}\frac{\partial g}{\partial x}=0 \qquad (1b)$$

$$\lambda y\frac{\partial g}{\partial y}+(f-x)\frac{\partial g}{\partial x}+\frac{\Gamma g}{1-Z_o h}\left(\frac{\partial f}{\partial x}+\frac{2f}{x}\right)-\lambda g=0 \qquad (1c)$$

where $x(r,t)=r/R(t)$, $y(t)=\left(a_o/U(t)\right)^2$, $u=U\,f(x,y)$, $p=\rho_o U^2 g(x,y)$, $\rho=\rho_o\,h(x,y)$, $U=dR/dt$, $\omega=t/t_o$, $\xi=R/R_o$ and $R_o=a_o\,t_o$. Here $r$ is the Eulerian coordinate measured from the centre of explosion and $t$ is the time co-ordinate measured from the instant of explosion. The position of shock front $R(t)$ measured from the centre of explosion is supposed to be monotonically increasing function of $t$. The variable $x$ represents the relative position with respect to the shock front. It varies from 0 at the centre to infinity and is equal to 1 at the shock position. The variable $y$ relates the inverse square of the shock velocity $U$ normalized by the initial sound speed $a_o$. Thus, the domain $(r,t)$ is transformed into the domain $(x,y)$. The quantities $f$, $g$, $h$, $\omega$ and $\xi$ are non-dimensional velocity of fluid, pressure, density, time-coordinate and field-coordinate, respectively. $R_o$ is a reference-front radius and it depends on the energy-input parameter $k$ and the pressure $p_o$ of undisturbed medium. The shock Mach number $M=1/\sqrt{y}$ is associated with the decay parameter $\lambda(y)$. The initial volume fraction of the solid particles $Z_o$ in the gas-particle two-phase medium is given by $Z_o=k_p/[G(1-k_p)+k_p]$, where $G$ is the volumetric parameter (the ratio of the density of solid particles to the initial density of gas) and $k_p$ is the mass concentration of solid particles in the mixture. The ratio of the specific heats of the mixture is $\Gamma=[\gamma(1-k_p)+k_p\beta_{sp}]/[1-k_p+k_p\beta_{sp}]$, where $\gamma=c_p/c_v$ is the ratio of specific heats of the gas, and $\beta_{sp}$ is the ratio of the specific heats of the solid particles. The volumetric fraction of solid particles in the mixture is $Z=Z_o\rho/\rho_o$ and the speed of sound in the unperturbed medium is $a_o=\sqrt{\Gamma p_o/(1-Z_o)\rho_o}$.

In present study, $x=0$ corresponds to the ground zero ($r=0$) and $x=1$ corresponds to the shock front ($r=R$). Since $U\to\infty$ as $t\to 0$ and $U\to a_o$ as $t\to\infty$; $y=0$ and $y=1$ correspond to $t=0$ and $t=\infty$, respectively. This transformation converts the region of the blast wave into a bounded rectangle $(0,1)\times(0,1)$. Thus, the boundary conditions given by Eqs. (31)–(33) in Ref. [11] can be written at the shock front $(x=1)$ as:

$$f(1,y)=\frac{2(1-Z_o)}{(\Gamma+1)}(1-y),\quad g(1,y)=\frac{2(1-Z_o)}{\Gamma+1}\left(1-\frac{\Gamma-1}{2\Gamma}y\right),\quad h(1,y)=\frac{\Gamma+1}{\Gamma-1+2Z_o}\left(1+\frac{2(1-Z_o)}{(\Gamma-1+2Z_o)}y\right)^{-1}.$$

(2a-c)

Using the boundary condition given by equation (21) in Ref. [18], we can write the pressure across the shock front as: $p(R,t)/p_o=2\Gamma/y(\Gamma+1)-(\Gamma-1)/(\Gamma+1)$, thus, $y\propto(p_o/p)$ at the shock front, the solution tends to the similarity solution for infinitely strong shock ($R\to 0$) as $y\to 0$. Also for very weak shock wave $R \to \infty$, $y$ tends to 1, i.e., $U \to a_o$. The non-dimensional discontinuity conditions (2a-c) are not much affected by the value of $y$, which varies only from 0 to 1, and we may expect a similar insensitivity to $y$ in energy integral equation also. Thus, the variable $y$ is expected to have little effect in the solutions.

**Power Series Solutions**

In this section, we presented the procedure of obtaining a power series in terms of a small parameter $y$, i.e. $M^{-2}$, where $M$ is the upstream shock Mach number. For strong shock waves, the shock velocity $U$ is large compared with $a_o$ and $y$ is considered to be small, thus the quantities $f$, $g$ and $h$ can be expanded in rapidly convergent series of power of $y$ as:

$$f(x, y) = f_0(x) + y f_1(x) + y^2 f_2(x) + o(y^3) \qquad (3a)$$

$$g(x, y) = g_0(x) + y g_1(x) + y^2 g_2(x) + o(y^3) \qquad (3b)$$

$$h(x, y) = h_0(x) + y h_1(x) + y^2 h_2(x) + o(y^3) \qquad (3c)$$

where $f_i$, $g_i$ and $h_i$ are functions of $x$ only. For strong shock waves the value of the co-ordinate $y$ is small; in fact, the case $y = 0$ represents an infinitely strong shock wave. In view of Freeman's model [12], the non-dimensional shock radius $\xi$ may be written as:

$$\xi = \xi_0 y^{1/\lambda_0} \left(1 + \xi_1 y + \xi_2 y^2 + \xi_3 y^3 + o(y^4)\right) \qquad (4)$$

where $\lambda_0 = 2(3 - k)/(2 + k)$ and $\xi_0, \xi_1, \xi_2, \ldots$ are constants. The velocity modulus $\omega$, is defined as $\omega = d \ln \xi / d \ln t = Ut/R$. Using $\omega = t/t_o$, $\xi = R/R_o$ and $R_o = a_o t_o$, we can write $\dfrac{d\xi}{d\omega} = \dfrac{t_o}{R_o} \dfrac{dR}{dt} = \dfrac{U}{a_o} = y^{-1/2}$ which yields $\omega = \int y^{1/2} \dfrac{d\xi}{dy} dy$. On integration, we obtain the velocity modulus $\omega$ as follows:

$$\omega = \omega_0 \xi_0 y^{1/\omega_0 \lambda_0} \left(1 + \omega_1 \xi_1 y + \omega_2 \xi_2 y^2 + \omega_3 \xi_3 y^3 + o(y^4)\right) \qquad (5)$$

where $\omega_0 = 2/(\lambda_0 + 2)$, $\omega_1 = (\lambda_0 + 1)(\lambda_0 + 2)/(3\lambda_0 + 2)$,

$\omega_2 = (\lambda_0 + 2)(2\lambda_0 + 1)/(5\lambda_0 + 2)$, $\omega_3 = (\lambda_0 + 2)(3\lambda_0 + 1)/(7\lambda_0 + 2)$.

Using relation $\lambda = d \ln y / d \ln \xi$ and Eq. (4), the shock decay parameter $\lambda(y)$ can be written as:

$$\lambda = \lambda_0 + \lambda_1 y + \lambda_2 y^2 + \lambda_3 y^3 + o(y^4) \qquad (6)$$

where $\lambda_1 = -\xi_1 \lambda_0^2$, $\lambda_2 = \xi_1^2 \lambda_0^2 - 2\xi_2 \lambda_0^2 + \xi_1^2 \lambda_0^3$, $\lambda_3 = -\xi_1^3 \lambda_0^2 + 3\xi_1 \xi_2 \lambda_0^2 - 3\xi_3 \lambda_0^2 - 2\xi_1^3 \lambda_0^3 + 4\xi_1 \xi_2 \lambda_0^3 - \xi_1^3 \lambda_0^4$,

Inserting Eqs. (3a-c) in the energy integral $J = \int_{x_P}^{1} \left( \dfrac{f^2 h}{2} + \dfrac{(1 - Z_o h) g}{(\Gamma - 1)} \right) x^2 dx$ (for detail see equation (48) in Ref. [11]), we have

$$J = J_0[1+\sigma_1 y+\sigma_2 y^2+\sigma_3 y^3+o(y^4)] \tag{7}$$

where $$J_0 = \int_{x_P}^{1}\left(\frac{f_0^2 h_0}{2}+\frac{(1-Z_o h_0)g_0}{\Gamma-1}\right)x^2 dx \tag{8a}$$

$$J_0\sigma_1 = \int_{x_P}^{1}\left(\frac{f_0(f_0 h_1+2f_1 h_0)}{2}+\frac{(1-Z_o h_0)g_1-Z_o g_0 h_1}{\Gamma-1}\right)x^2 dx \tag{8b}$$

$$J_0\sigma_2 = \int_{x_P}^{1}\left(\frac{f_0(f_0 h_2+2f_2 h_0)+f_1(f_1 h_0+2f_0 h_1)}{2}+\frac{(1-Z_o h_0)g_2-Z_o(g_0 h_2+g_1 h_1)}{\Gamma-1}\right)x^2 dx \tag{8c}$$

The non-dimensional energy integral as given by equation (39) in Ref. [11] can be written for spherical shock wave as:

$$J = \frac{y(1-Z_o)}{\Gamma}\left(\frac{\omega^k}{\xi^3}+\frac{(1-Z_o)}{3(\Gamma-1)}\right) \tag{9}$$

Substituting Eqs. (4)–(7) into (9), ensures its power series form provided $1/\lambda_0$ and $(\lambda_0+2)/2\lambda_0$ are positive integers, i.e., $1/\lambda_0 = I$ (where $I$ is a positive integer), which is equivalent to $k = 2\{(I(\alpha+1)-1)/(2I+1)\}$; where $\alpha = 0, 1 \text{ and } 2$ for plane, cylindrical and spherical symmetry of the shock, respectively. It can be seen that for these permissible values of $k$, the term $k(\lambda_0+2)/2\lambda_0$ automatically becomes a positive integer. The first permissible value of $k$ is $4/3$ for spherical symmetry of shock.

Now, substituting Eqs. (3a-c) and (6) in Eqs. (1a-c) and comparing the coefficients of the same powers of $y$ on both sides of (1a-c), we get the following system of equations:

For zeroth power of $y$,

$$(f_0-x)h_0^{/}+\frac{2f_0 h_0}{x}+h_0 f_0^{/}=0 \tag{10a}$$

$$(f_0-x)h_0 f_0^{/}-\frac{f_0 h_0}{2}+g_0^{/}=0 \tag{10b}$$

$$(f_0-x)(1-Z_o h_0)g_0^{/}-(1-Z_o h_0)g_0+\Gamma g_0\left(\frac{2f_0}{x}+f_0^{/}\right)=0 \tag{10c}$$

For first power of $y$,

$$\frac{2f_1 h_0}{x}+h_1+\frac{2f_0 h_1}{x}+h_1 f_0^{/}+h_0 f_1^{/}+f_1 h_0^{/}-xh_1^{/}+f_0 h_1^{/}=0 \tag{11a}$$

$$\frac{\xi_1 f_0 h_0}{2}+\frac{f_1 h_0}{2}-\frac{f_0 h_1}{2}+f_1 h_0 f_0^{/}+(f_0-x)\left(h_1 f_0^{/}+h_0 f_1^{/}\right)+g_1^{/}=0 \tag{11b}$$

$$\begin{aligned}&\xi_1 g_0(1-Z_o h_0)+Z_o g_0 h_1+\Gamma g_1\left(\frac{2f_0}{x}+f_0^{/}\right)+\Gamma g_0\left(\frac{2f_1}{x}+f_1^{/}\right)+\\&f_1(1-Z_o h_0)g_0^{/}-(f_0-x)Z_o h_1 g_0^{/}+(f_0-x)(1-Z_o h_0)g_1^{/}=0\end{aligned} \tag{11c}$$

For second power of $y$,

$$\frac{2f_2h_0}{x}-\xi_1h_1+\frac{2f_1h_1}{x}+2h_2+\frac{2f_0h_2}{x}+h_2f_0^{/}+h_1f_1^{/}+h_0f_2^{/}+f_2h_0^{/}+f_1h_1^{/}+\left(f_0-x\right)h_2^{/}=0 \tag{12a}$$

$$\begin{gathered}\left(\xi_2-\xi_1^2\right)f_0h_0-\xi_1f_1h_0+2f_2h_0+f_1h_1+\frac{\xi_1}{2}\left(f_1h_0+f_0h_1\right)-\frac{1}{2}\left(f_2h_0+f_1h_1+f_0h_2\right)+\\ f_2h_0f_0^{/}+f_1\left(h_1f_0^{/}+h_0f_1^{/}\right)+\left(f_0-x\right)\left(h_2f_0^{/}+h_1f_1^{/}+h_0f_2^{/}\right)+g_2^{/}=0\end{gathered} \tag{12b}$$

$$\begin{gathered}\left(\xi_2-\xi_1^2\right)g_0\left(1-Z_oh_0\right)-\xi_1g_1\left(1-Z_oh_0\right)+g_2\left(1-Z_oh_0\right)+\xi_1g_1\left(1-Z_oh_0\right)-\xi_1Z_og_0h_1+Z_og_0h_2+\\ \Gamma g_2\left(\frac{2f_0}{x}+f_0^{/}\right)+\Gamma g_1\left(\frac{2f_1}{x}+f_1^{/}\right)+\Gamma g_0\left(\frac{2f_2}{x}+f_2^{/}\right)+f_2\left(1-Z_oh_0\right)g_0^{/}-f_1Z_oh_1g_0^{/}+\\ f_1\left(1-Z_oh_0\right)g_1^{/}-\left(f_0-x\right)Z_o\left(h_2g_0^{/}+h_1g_1^{/}+h_0g_2^{/}\right)+\left(f_0-x\right)g_2^{/}=0\end{gathered} \tag{12c}$$

where prime denotes derivatives with respect to $x$. The boundary conditions at the shock front $x=1$ can be determined by substituting Eqs. (3a-c) in Eqs. (2a-c), we have boundary conditions respectively for differential equations (10a-c), (11a-c) and (12a-c), in a similar manner,

$$f_0(1)=2(1-Z_o)/(1+\Gamma)\,,\ \ g_0(1)=2(1-Z_o)/(1+\Gamma)\,,\ h_0(1)=(1+\Gamma)/(\Gamma-1+2Z_o) \tag{13a-c}$$

$$f_1(1)=-2(1-Z_o)/(1+\Gamma)\,,\ \ g_1(1)=-(\Gamma-1)(1-Z_o)/\Gamma(1+\Gamma)\,,\ \ h_1(1)=0 \tag{14a-c}$$

$$f_2(1)=0\,,\ \ g_2(1)=0\,,\ \ h_2(1)=0 \tag{15a-c}$$

The zeroth order shock boundary conditions (13a-c) are simply those for an infinitely strong shock wave of limiting density ratio. Thus, the solution of the zeroth order equations (10a-c) with the corresponding boundary conditions (13a-c) yields the solution for an infinitely strong shock wave. Hence, for very large values of the initiation energy $E_o$, the first and higher order conditions become zero and the shock remains strong throughout under this condition.

Since the higher order equations have the same determinant as the zeroth order, there are no singularities involved in the solutions. The zeroth order equations (10a-c) with boundary conditions (13a-c) can be integrated numerically using the Runge-Kutta method of fourth order. The value of $J_0$ can be computed from Eq. (8a) using numerical solution of zeroth order equations. Now, $\xi_0$ can be calculated from Eq. (9) using Eqs. (3a-c), (4) and known value of $J_0$. However, the first and second order equations (11a-c, 12a-c) cannot be integrated directly because of the presence of the constants $\xi_1$ and $\xi_2$. To obtain the solutions for these first and second order equations, we first write the functions as:

$$f_1(x)=f_{11}(x)+\xi_1f_{12}(x)\,,\ g_1(x)=g_{11}(x)+\xi_1g_{12}(x)\,,\ \ h_1(x)=h_{11}(x)+\xi_1h_{12}(x)\,,\ \ \sigma_1=\sigma_{11}+\xi_1\sigma_{12} \tag{16a-d}$$

$$f_2(x)=f_{21}(x)+\xi_2f_{22}(x)\,,\ g_2(x)=g_{21}(x)+\xi_2g_{22}(x)\,,\ h_2(x)=h_{21}(x)+\xi_2h_{22}(x)\,,\ \sigma_2=\sigma_{21}+\xi_2\sigma_{22} \tag{17a-d}$$

In these, $\xi_1$ can be eliminated by substituting Eqs. (16a-c) into Eqs. (11a-c) and grouping the terms with and without $\xi_1$. This leads, after some manipulation, to the following two pairs of coupled ordinary first order equations with the corresponding boundary conditions:

$$\frac{2f_{11}h_0}{x}+h_{11}+\frac{2f_0h_{11}}{x}+h_{11}f_0^{/}+h_0f_{11}^{/}+f_{11}h_0^{/}-xh_{11}^{/}+f_0h_{11}^{/}=0 \tag{18a}$$

$$\frac{f_{11}h_0}{2}-\frac{f_0h_{11}}{2}+f_{11}h_0f_0'-xh_{11}f_0'+f_0h_{11}f_0'-xh_0f_{11}'+f_0h_0f_{11}'+g_{11}'=0 \qquad (18b)$$

$$\frac{2\Gamma f_{11}g_0}{x}+\frac{2\Gamma f_0g_{11}}{x}+Z_og_0h_{11}+\Gamma g_{11}f_0'+\Gamma g_0f_{11}'+f_{11}g_0'-Z_of_{11}h_0g_0'+$$
$$xZ_oh_{11}g_0'-Z_of_0h_{11}g_0'-xg_{11}'+f_0g_{11}'+xZ_oh_0g_{11}'-Z_of_0h_0g_{11}'=0 \qquad (18c)$$

$$f_{11}(1)=-2(1-Z_o)/(1+\Gamma)\,,\;\; g_{11}(1)=-(\Gamma-1)(1-Z_o)/\Gamma(1+\Gamma)\,,\;\; h_{11}(1)=0 \qquad (19a\text{-}c)$$

$$\frac{2f_{12}h_0}{x}+h_{12}+\frac{2f_0h_{12}}{x}+h_{12}f_0'+h_0f_{12}'+f_{12}h_0'-xh_{12}'+f_0h_{12}'=0 \qquad (20a)$$

$$\frac{f_0h_0}{2}+\frac{f_{12}h_0}{2}-\frac{f_0h_{12}}{2}+f_{12}h_0f_0'-xh_{12}f_0'+f_0h_{12}f_0'-xh_0f_{12}'+f_0h_0f_{12}'+g_{12}'=0 \qquad (20b)$$

$$g_0+\frac{2\Gamma f_{12}g_0}{x}+\frac{2\Gamma f_0g_{12}}{x}-Z_og_0h_0+Z_og_0h_{12}+\Gamma g_{12}f_0'+\Gamma g_0f_{12}'+f_{12}g_0'-$$
$$Z_of_{12}h_0g_0'+xZ_oh_{12}g_0'-Z_of_0h_{12}g_0'-xg_{12}'+f_0g_{12}'+xZ_oh_0g_{12}'-Z_of_0h_0g_{12}'=0 \qquad (20c)$$

$$f_{12}(1)=0\,,\;\; g_{12}(1)=0\,,\;\; h_{12}(1)=0 \qquad (21a\text{-}c)$$

Numerical values of $\sigma_{11}$ and $\sigma_{12}$ are obtained from Eq. (9) using Eqs. (3a-c), (4) and known values of $J_0$, and $\xi_0$. The value of $\sigma_1$ is calculated from Eq. (8b) and thus, $\xi_1$ is obtained using Eq. (16d).

By the same procedure $\xi_2$ can be eliminated by substituting Eqs. (17a-c) into Eqs. (12a-c), which leads to the following two pairs of coupled ordinary second order equations with the corresponding boundary conditions:

$$\frac{2f_{21}h_0}{x}-\xi_1h_{11}+\frac{2f_{11}h_{11}}{x}+\frac{2\xi_1f_{12}h_{11}}{x}-\xi_1^2h_{12}+\frac{2\xi_1f_{11}h_{12}}{x}+\frac{2\xi_1^2f_{12}h_{12}}{x}+2h_{21}+$$
$$\frac{2f_0h_{21}}{x}+h_{21}f_0'+h_{11}f_{11}'+\xi_1h_{12}f_{11}'+\xi_1h_{11}f_{12}'+\xi_1^2h_{12}f_{12}'+h_0f_{21}'+f_{21}h_0'+f_{11}h_{11}'+ \qquad (22a)$$
$$\xi_1f_{12}h_{11}'+\xi_1f_{11}h_{12}'+\xi_1^2f_{12}h_{12}'-xh_{21}'+f_0h_{21}'=0$$

$$-\xi_1^2f_0h_0-\frac{\xi_1f_{11}h_0}{2}-\frac{\xi_1^2f_{12}h_0}{2}+\frac{3f_{21}h_0}{2}+\frac{\xi_1f_0h_{11}}{2}+\frac{f_{11}h_{11}}{2}+\frac{\xi_1f_{12}h_{11}}{2}+\frac{\xi_1^2f_0h_{12}}{2}+\frac{\xi_1f_{11}h_{12}}{2}+$$
$$\frac{\xi_1^2f_{12}h_{12}}{2}-\frac{f_0h_{21}}{2}+f_{21}h_0f_0'+f_{11}h_{11}f_0'+\xi_1f_{12}h_{11}f_0'+\xi_1f_{11}h_{12}f_0'+\xi_1^2f_{12}h_{12}f_0'-xh_{21}f_0'+ \qquad (22b)$$
$$f_0h_{21}f_0'+f_{11}h_0f_{11}'+\xi_1f_{12}h_0f_{11}'-xh_{11}f_{11}'+f_0h_{11}f_{11}'-x\xi_1h_{12}f_{11}'+\xi_1f_0h_{12}f_{11}'+\xi_1f_{11}h_0f_{12}'+$$
$$\xi_1^2f_{12}h_0f_{12}'-x\xi_1h_{11}f_{12}'+\xi_1f_0h_{11}f_{12}'-x\xi_1^2h_{12}f_{12}'+\xi_1^2f_0h_{12}f_{12}'-xh_0f_{21}'+f_0h_0f_{21}'+g_{21}'=0$$

$$
\begin{aligned}
&-2\xi_1^2 g_0 + \frac{2\Gamma f_{21} g_0}{x} + \frac{2\Gamma f_{11} g_{11}}{x} + \frac{2\Gamma \xi_1 f_{11} g_{11}}{x} + \frac{2\Gamma \xi_1 f_{11} g_{12}}{x} + \frac{2\Gamma \xi_1^2 f_{12} g_{12}}{x} + g_{21} + \\
&\frac{2\Gamma f_0 g_{21}}{x} + 2Z_o \xi_1^2 g_0 h_0 - Z_o g_{21} h_0 - Z_o \xi_1 g_0 h_{11} - Z_o \xi_1^2 g_0 h_{12} + Z_o g_0 h_{21} + \Gamma g_{21} f_0^{/} + \\
&\Gamma g_{11} f_{11}^{/} + \Gamma \xi_1 g_{12} f_{11}^{/} + \Gamma \xi_1 g_{11} f_{12}^{/} + \Gamma \xi_1^2 g_{12} f_{12}^{/} + \Gamma g_0 f_{21}^{/} + f_{21} g_0^{/} - Z_o f_{21} h_0 g_0^{/} - \\
&Z_o f_{11} h_{11} g_0^{/} - Z_o \xi_1 f_{12} h_{11} g_0^{/} - Z_o \xi_1 f_{11} h_{12} g_0^{/} - z_o \xi_1^2 f_{12} h_{12} g_0^{/} + x Z_o h_{21} g_0^{/} - Z_o f_0 h_{21} g_0^{/} + \\
&f_{11} g_{11}^{/} + \xi_1 f_{12} g_{11}^{/} - Z_o f_{11} h_0 g_{11}^{/} - Z_o \xi_1 f_{12} h_0 g_{11}^{/} + x Z_o h_{11} g_{11}^{/} - Z_o f_0 h_{11} g_{11}^{/} + x Z_o \xi_1 h_{12} g_{11}^{/} - \\
&Z_o \xi_1 f_0 h_{12} g_{11}^{/} + \xi_1 f_{11} g_{12}^{/} + \xi_1^2 f_{12} g_{12}^{/} - Z_o \xi_1 f_{11} h_0 g_{12}^{/} - Z_o \xi_1^2 f_{12} h_0 g_{12}^{/} + x Z_o \xi_1 h_{11} g_{12}^{/} - \\
&Z_o \xi_1 f_0 h_{11} g_{12}^{/} + x Z_o \xi_1^2 h_{12} g_{12}^{/} - Z_o \xi_1^2 f_0 h_{12} g_{12}^{/} - x g_{21}^{/} + f_0 g_{21}^{/} + x Z_o h_0 g_{21}^{/} - Z_o f_0 h_0 g_{21}^{/} = 0
\end{aligned}
\tag{22c}
$$

$$f_{21}(1) = 0\,,\ \ h_{21}(1) = 0\,,\ \ g_{21}(1) = 0 \tag{23a-c}$$

$$\frac{2 f_{22} h_0}{x} + 2h_{22} + \frac{2 f_0 h_{22}}{x} + h_{22} f_0^{/} + h_0 f_{22}^{/} + f_{22} h_0^{/} - x h_{22}^{/} + f_0 h_{22}^{/} = 0 \tag{24a}$$

$$f_0 h_0 + \frac{3 f_{22} h_0}{2} - \frac{f_0 h_{22}}{2} + f_{22} h_0 f_0^{/} - x h_{22} f_0^{/} + f_0 h_{22} f_0^{/} - x h_0 f_{22}^{/} + f_0 h_0 f_{22}^{/} + g_{22}^{/} = 0 \tag{24b}$$

$$
\begin{aligned}
&2g_0 + \frac{2\Gamma f_{22} g_0}{x} + g_{22} + \frac{2\Gamma f_0 g_{22}}{x} - 2Z_o g_0 h_0 - Z_o g_{22} h_0 + Z_o g_0 h_{22} + \Gamma g_{22} f_0^{/} + \Gamma g_0 f_{22}^{/} + \\
&f_{22} g_0^{/} - Z_o f_{22} h_0 g_0^{/} + x Z_o h_{22} g_0^{/} - Z_o f_0 h_{22} g_0^{/} - x g_{22}^{/} + f_0 g_{22}^{/} + x Z_o h_0 g_{22}^{/} - Z_o f_0 h_0 g_{22}^{/} = 0
\end{aligned}
\tag{24c}
$$

$$f_{22}(1) = 0\,,\ \ h_{22}(1) = 0\,,\ \ g_{22}(1) = 0 \tag{25a-c}$$

Numerical values of $\sigma_{21}$ and $\sigma_{22}$ are obtained from Eq. (9) using Eqs. (3a-c), (4) and known values of $J_0$, $\xi_0$ and $\xi_1$. The value of $\sigma_2$ is calculated from Eq. (8c) and thus, $\xi_2$ is obtained using Eq. (17d). The solutions for the third and higher orders are found in a similar manner and thus, $f_n$, $g_n$, $h_n$, $\xi_n$ for all $n$ are to be found successively. Finally, the non-dimensional expressions for the distribution of the velocity of fluid, the pressure, the density, the speed of sound, the adiabatic compressibility and the change-in-entropy behind the strong spherical shock front can be, respectively, written as:

$$\frac{u}{a_o} = \frac{1}{\sqrt{y}} \sum_{n=0}^{\infty} f_n(x) y^n \tag{26a}$$

$$\frac{p}{p_o} = \frac{\Gamma}{(1 - Z_o) y} \sum_{n=0}^{\infty} g_n(x) y^n \tag{26b}$$

$$\frac{\rho}{\rho_o} = \sum_{n=0}^{\infty} h_n(x) y^n \tag{26c}$$

$$\frac{a}{a_o} = \left( \Gamma \sum_{n=0}^{\infty} g_n(x) y^n \Big/ (1 - Z) y \sum_{n=0}^{\infty} h_n(x) y^n \right)^{1/2} \tag{26d}$$

$$\tau(p_o) = (1 - Z)(1 - Z_o) y \Big/ \Gamma^2 \sum_{n=0}^{\infty} g_n(x) y^n \tag{26e}$$

$$\frac{\Delta s}{R_i}=\frac{(1-k_p)}{\Gamma-1}\ln\left(\frac{\Gamma}{(1-Z_o)y}\sum_{n=0}^{\infty}g_n(x)y^n\right)-\frac{\Gamma(1-k_p)}{\Gamma-1}\ln\left(\sum_{n=0}^{\infty}h_n(x)y^n\right)+\frac{\Gamma(1-k_p)}{\Gamma-1}\ln\left(\frac{1-Z}{1-Z_o}\right) \qquad (26f)$$

where $R_i$ is the specific gas constant of the dust-free gas.

## Results and Discussion

The distributions of flow quantities between the spherical shock front ( $x=1$ ) and the inner expanding surface or piston ( $x=x_p$ ) are obtained by the numerical integration of the set of differential Eqs. (10a-c), (18a-c), (20a-c), (22a-c) and (24a-c) with their corresponding boundary conditions (13a-c), (19a-c), (21a-c), (23a-c) and (25a-c) by the Runge–Kutta method of fourth order. The typical values of parameters are taken as $k=4/3$, $\gamma=7/5$, $M=5$, $\beta_{sp}=1$, $k_p=$ 0, 0.2, 0.4 and $G=$1, 10, 100 for numerical computation of flow-field quantities (26a-f) using Mathematica8. The parameter $k_p=0$ corresponds to the perfect gas. Also, the parameter $G=1$ corresponds to the case when the initial volume fraction of solid particles $Z_o$ in the mixture is equal to the mass fraction of solid particles $k_p$. In our analysis, we have assumed $Z_o$ to be a small constant. The parameter $k_p=$ 0.2, 0.4 with the parameter $G=$1, 10, 100 give small values of $Z_o$ [18]. The position of piston is determined by the kinematic condition $f(x_p)=x_p$, which states that the velocity of the fluid at the piston is equal to the position of the piston. Starting from the shock front, the numerical integration is carried out until the singularity of the solution $f(x_p)=x_p$ is reached. This marks the position of piston, i.e., the inner expanding surface $x_p$. Table 1 shows the position of the inner boundary surface $x_p$ and the values of $J_0$, $\xi_0$, $\xi_1$, $\xi_2$, $\sigma_1$ and $\sigma_2$ for $k=4/3$, $M=5$, $\gamma=7/5$, $\beta_{sp}=1$, $k_p=$ 0, 0.2, 0.4 and $G=$1, 10, 100. The distributions of the velocity of fluid $u/a_o$, the pressure $p/p_o$, the density $\rho/\rho_o$, the speed of sound $a/a_o$, the adiabatic compressibility $\tau(p_o)$ and the change-in-entropy $\Delta s/R_i$ in the region behind the spherical shock front with the propagation distance $x$ and the reciprocal square of Mach number $y$ are shown on Figures 1(a-f) and 2(a-f), respectively.

Figure 1. illustrates that the velocity of fluid, the pressure for $G=1$, the speed of sound for $G\geq 10$, the adiabatic compressibility for $G\geq 10$ and the change-in-entropy increase as we move towards the piston from the shock front, however, the pressure for $G\geq 10$ and the density for $G\geq 10$ show reverse trends. The density, the speed of sound and the compressibility remain almost unchanged for $G=1$ in the region between the shock front and the inner expanding surface (see Figs. 1c–e). Figure 2. shows that the velocity of fluid, the pressure, the speed of sound and the change-in-entropy decrease, however, the density and the adiabatic compressibility increase with increase in $y$. The variations of the velocity of fluid, the speed of sound and the change-in-entropy, in particular for $k_p=0.4$, $G=1$ differs greatly from the ideal gas (see

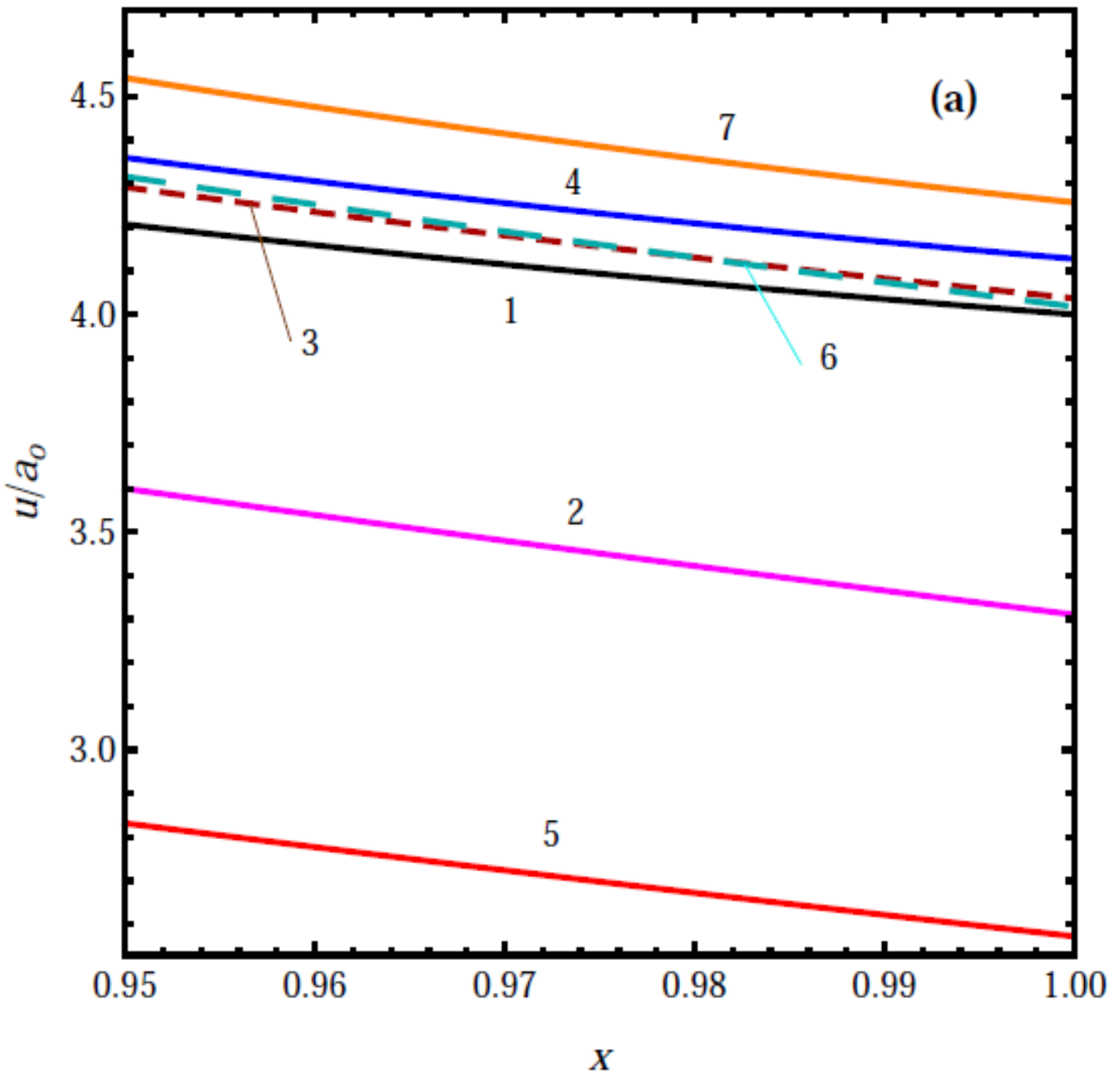

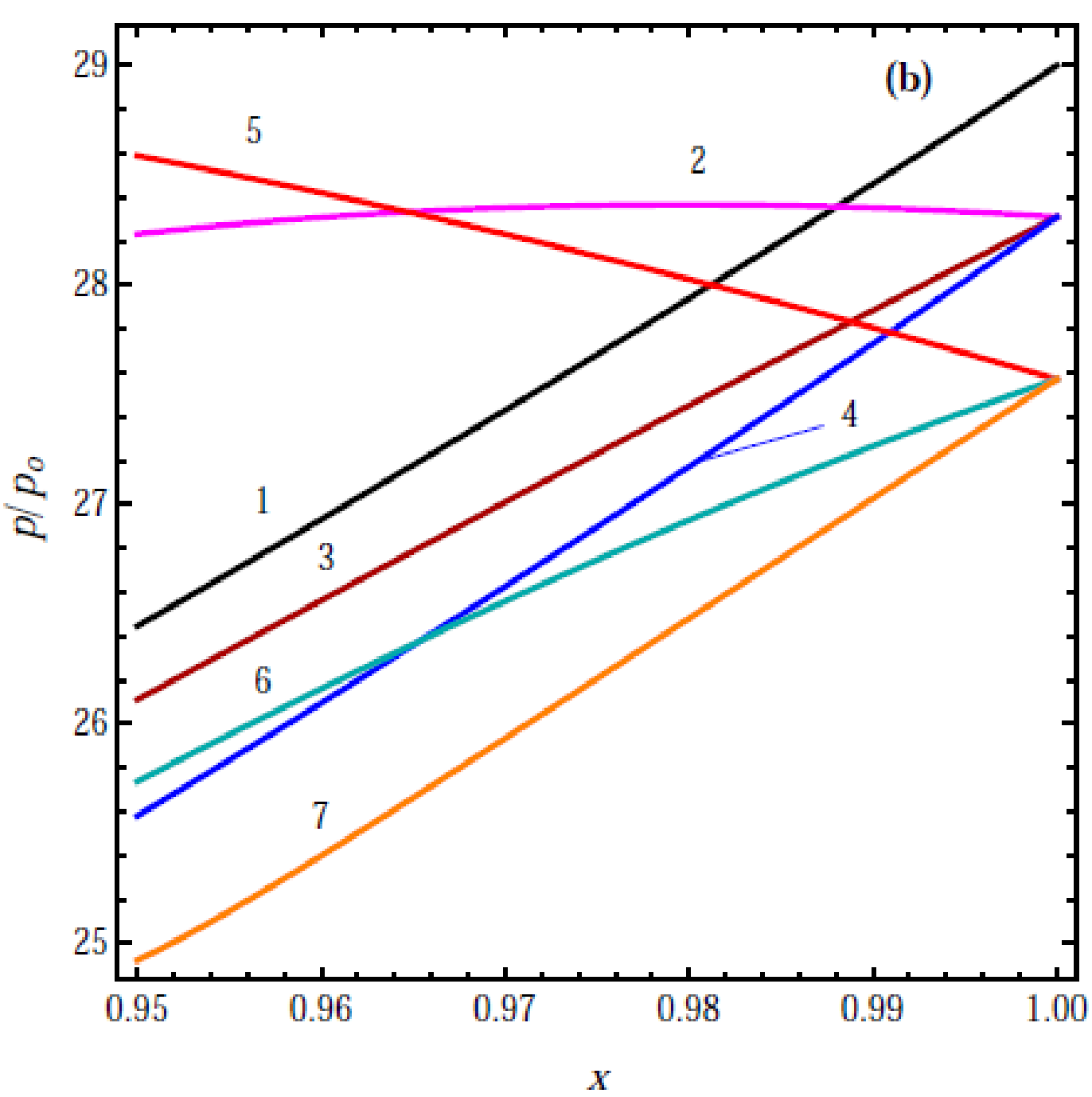

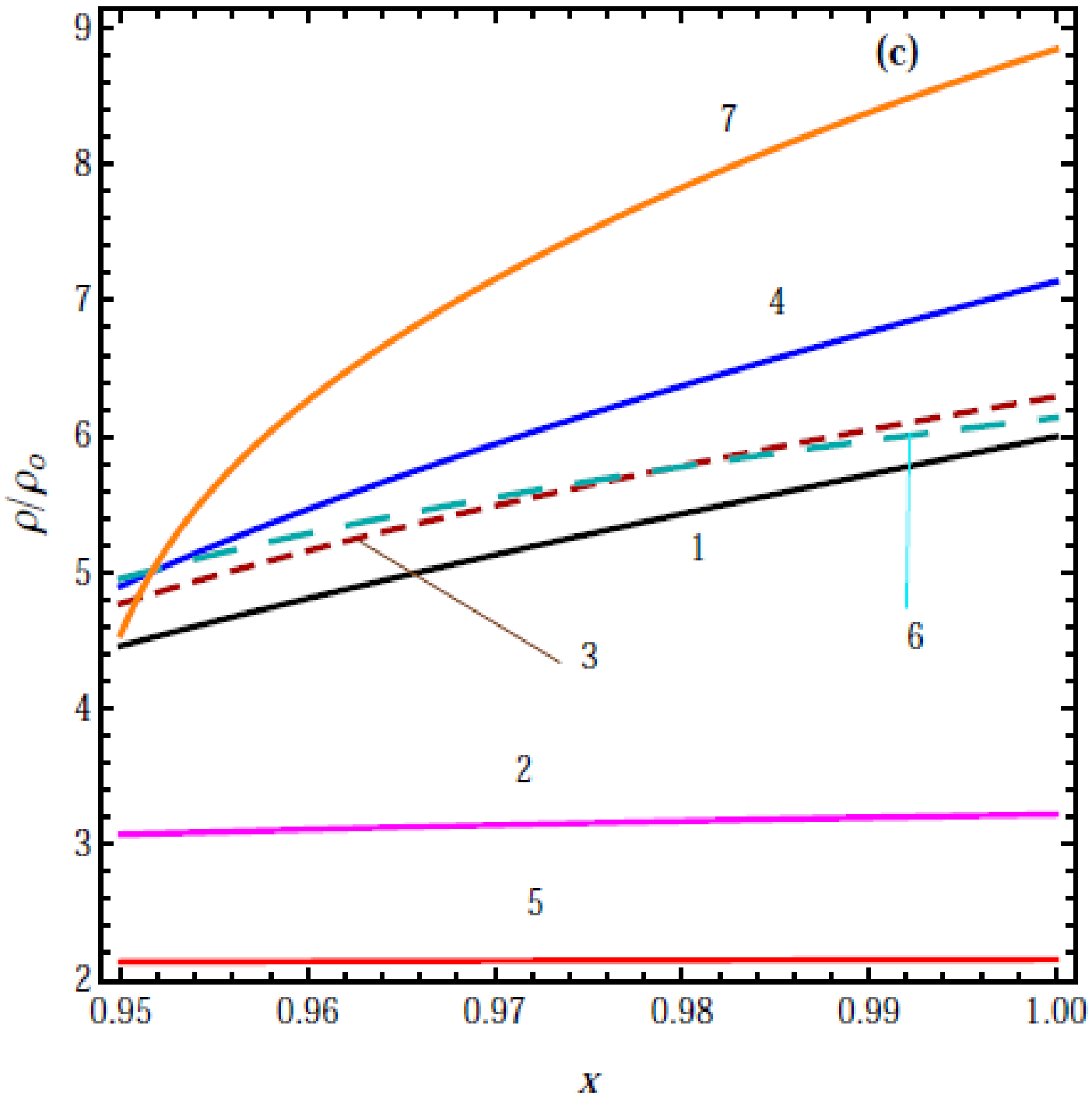

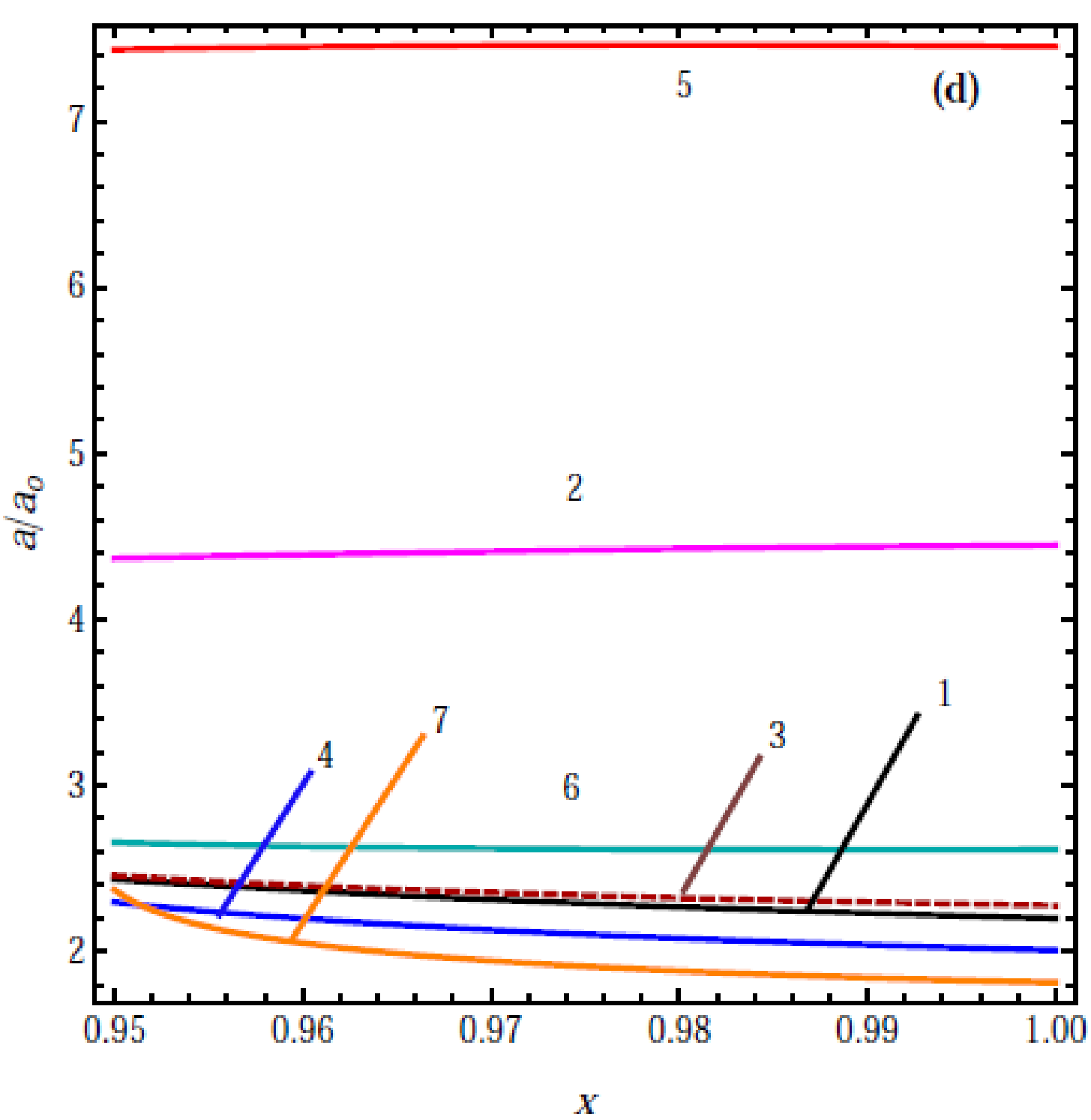

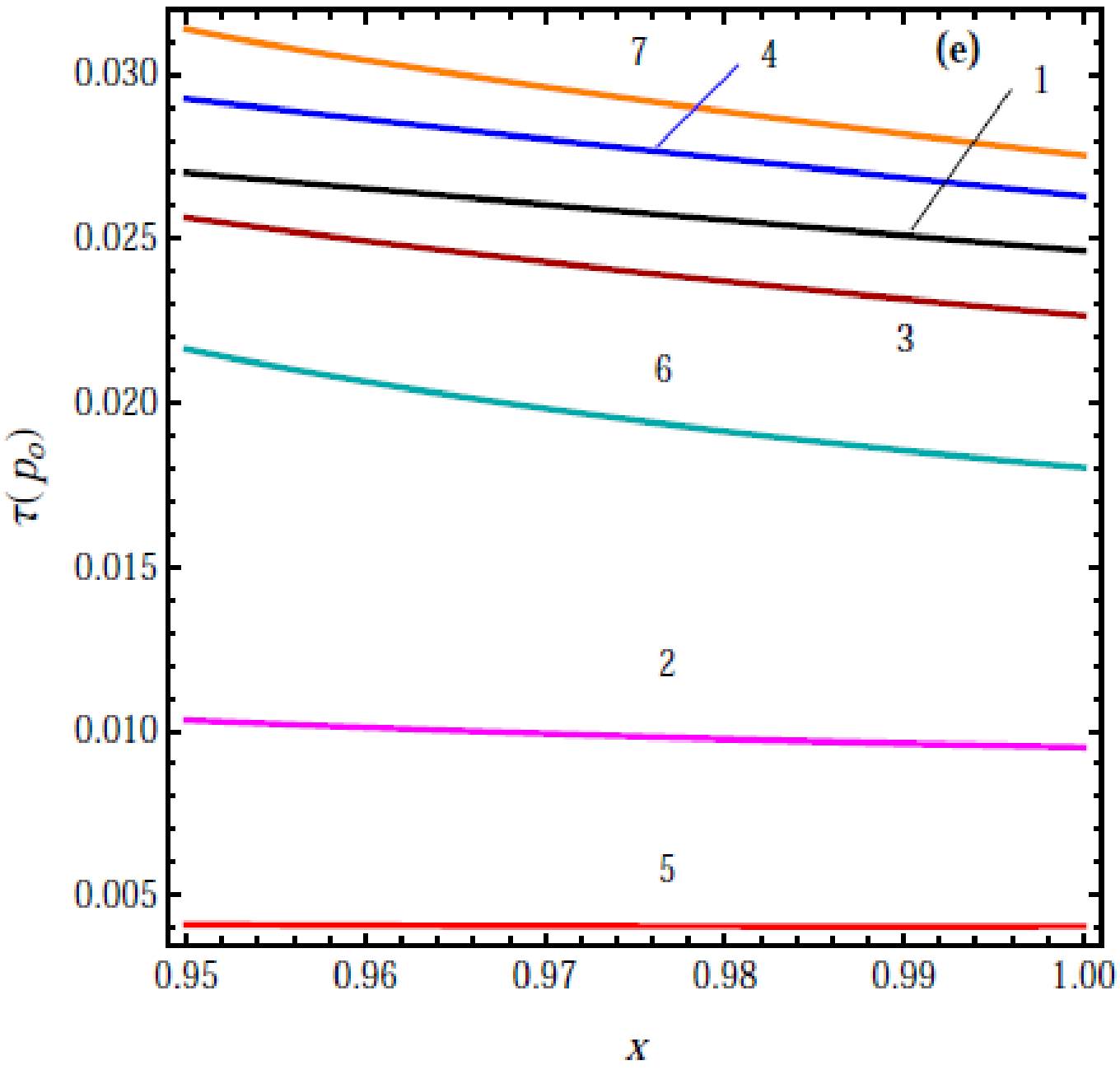

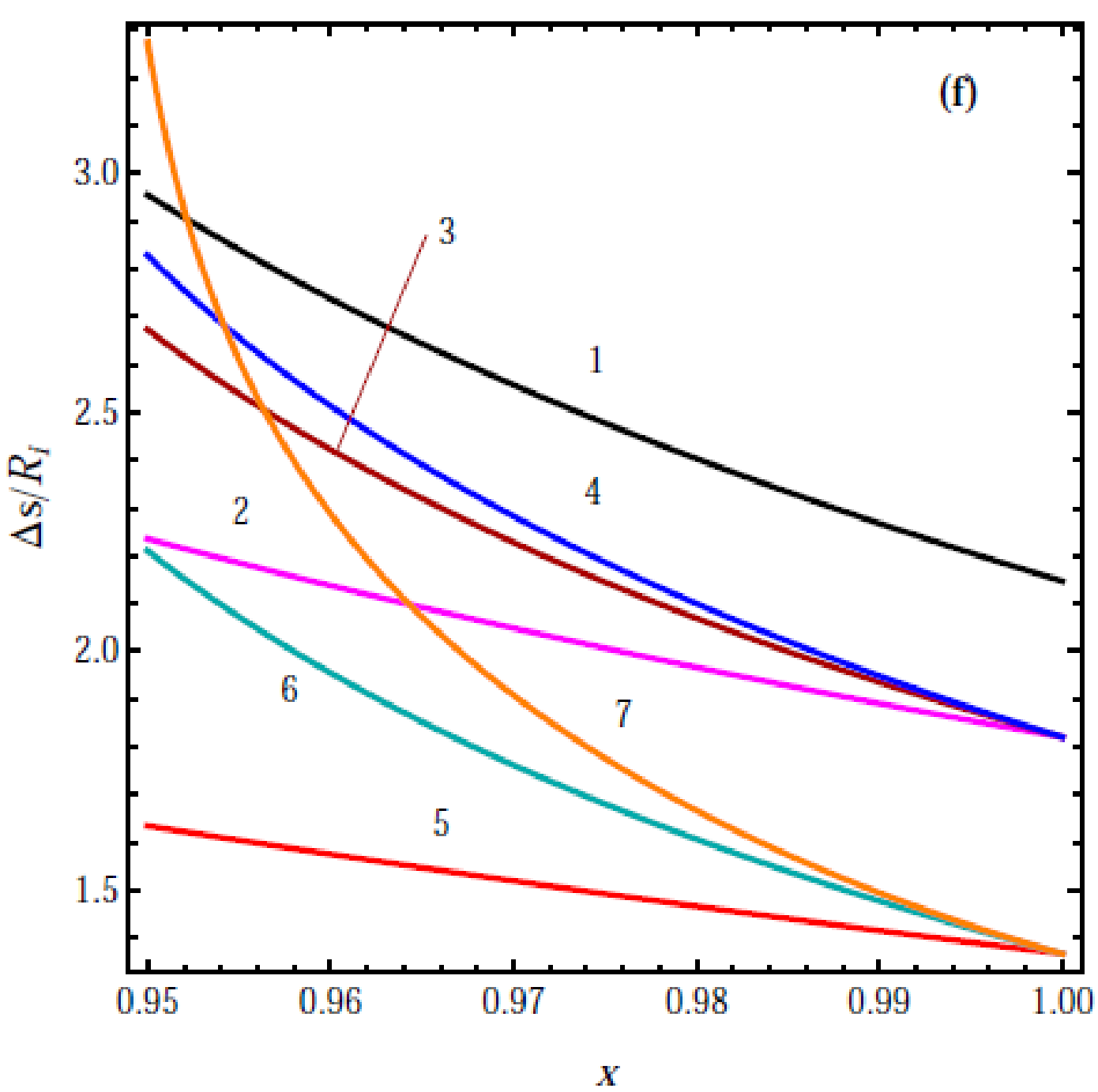

**Fig. 1** Variations of non-dimensional (a) velocity of fluid $u/a_o$, (b) pressure $p/p_o$, (c) density $\rho/\rho_o$, (d) speed of sound $a/a_o$, (e) adiabatic compressibility $\tau(p_o)$ and (f) change-in-entropy $\Delta s/R_i$ just behind the spherical shock front with the propagation distance $x$ for $\beta_{sp}=1$, $\gamma=7/5$, $y=0.04$ and various values of $k_p$ and $G$. **1**: $k_p=0$; **2**: $k_p=0.2, G=1$; **3**: $k_p=0.2, G=10$; **4**: $k_p=0.2, G=100$; **5**: $k_p=0.4, G=1$; **6**: $k_p=0.4, G=10$; **7**: $k_p=0.4, G=100$.

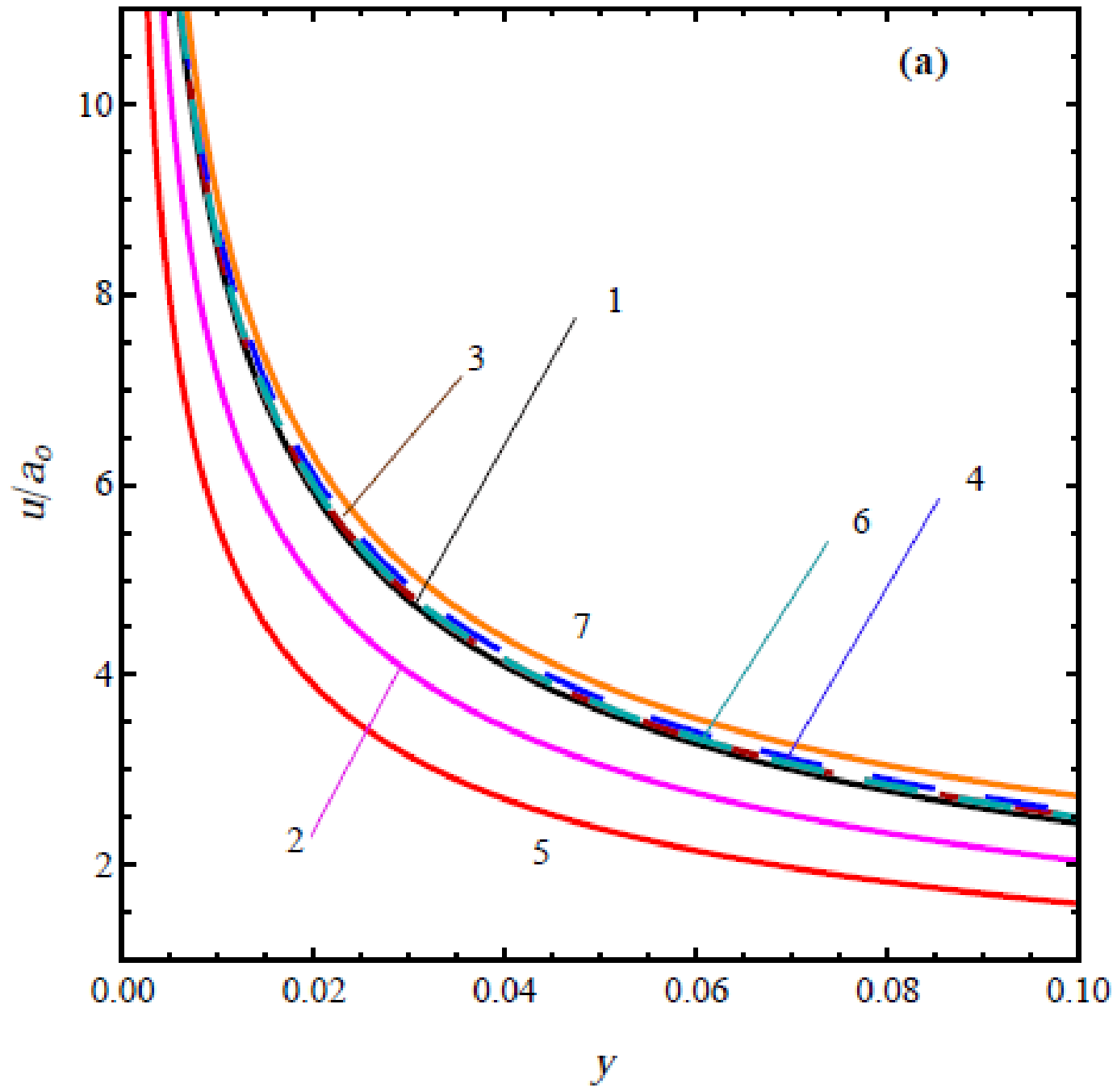

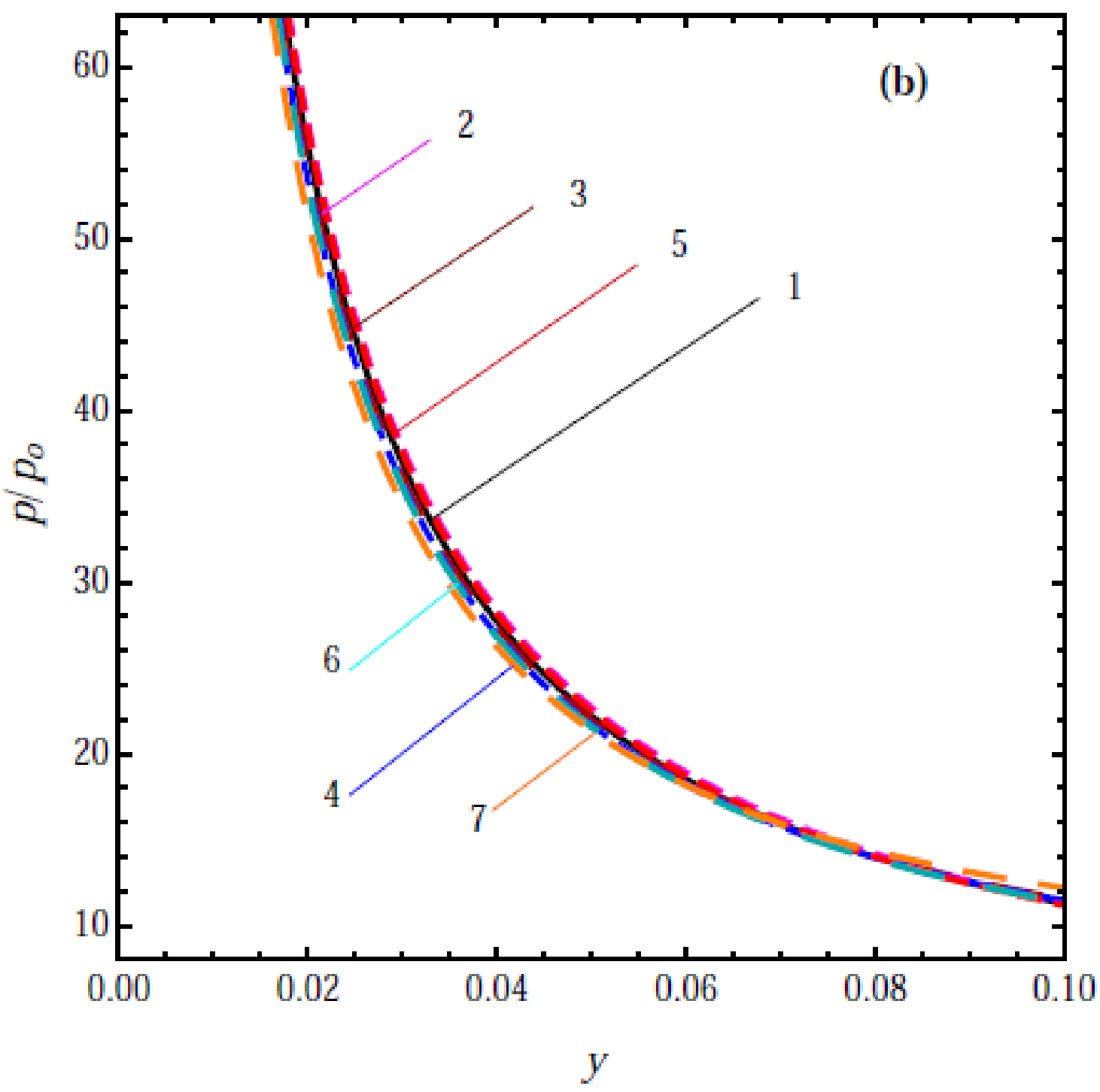

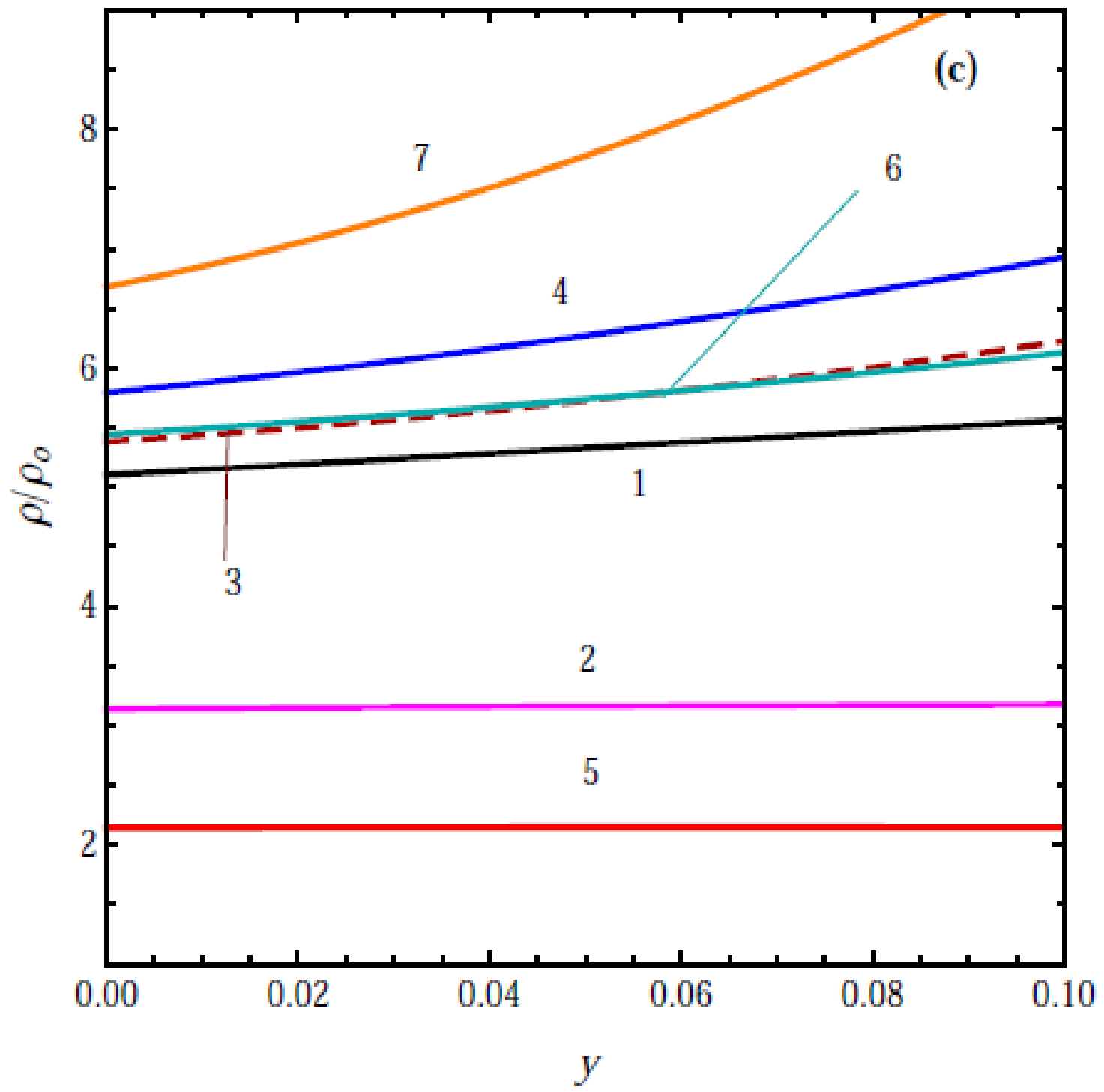

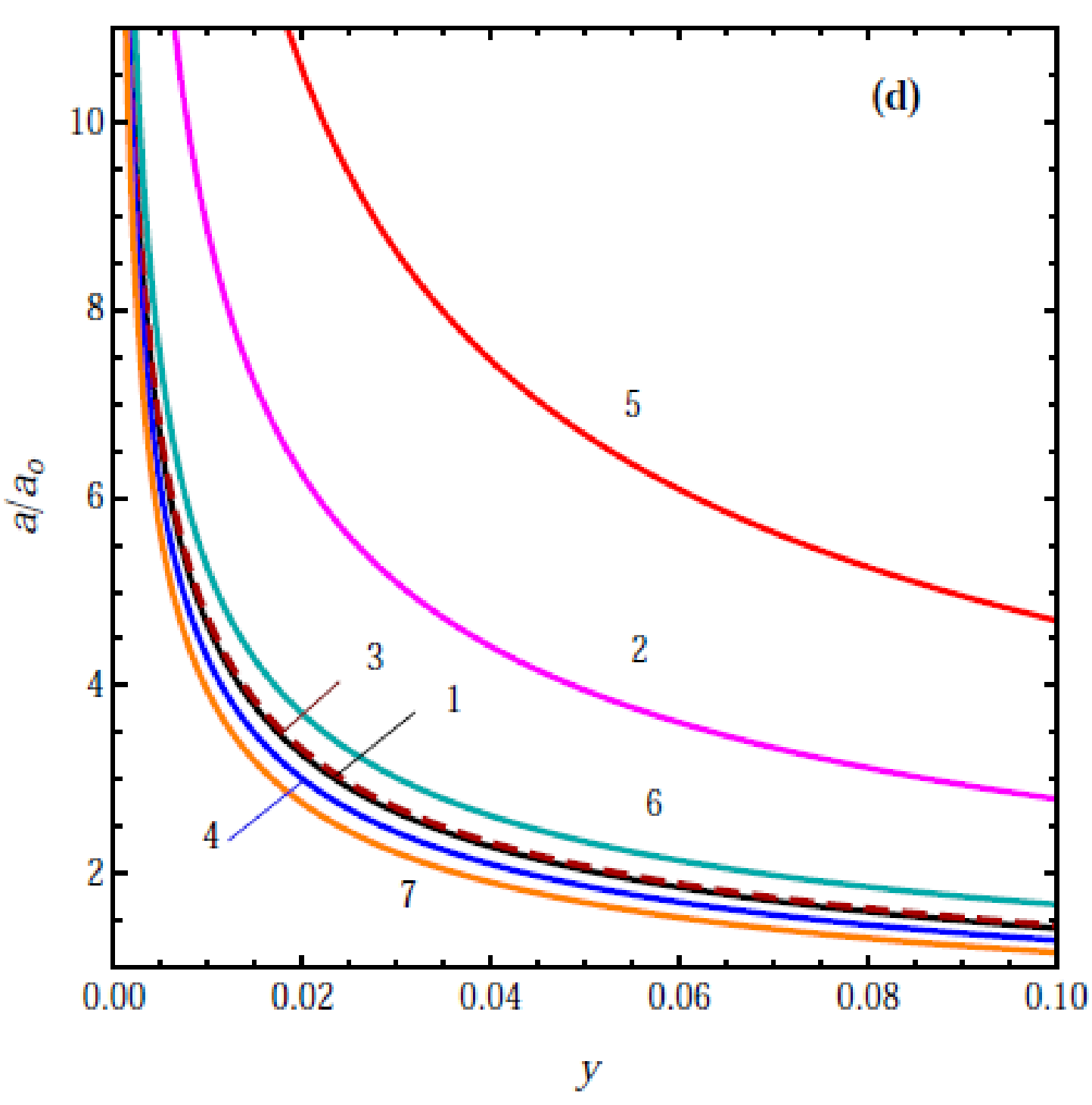

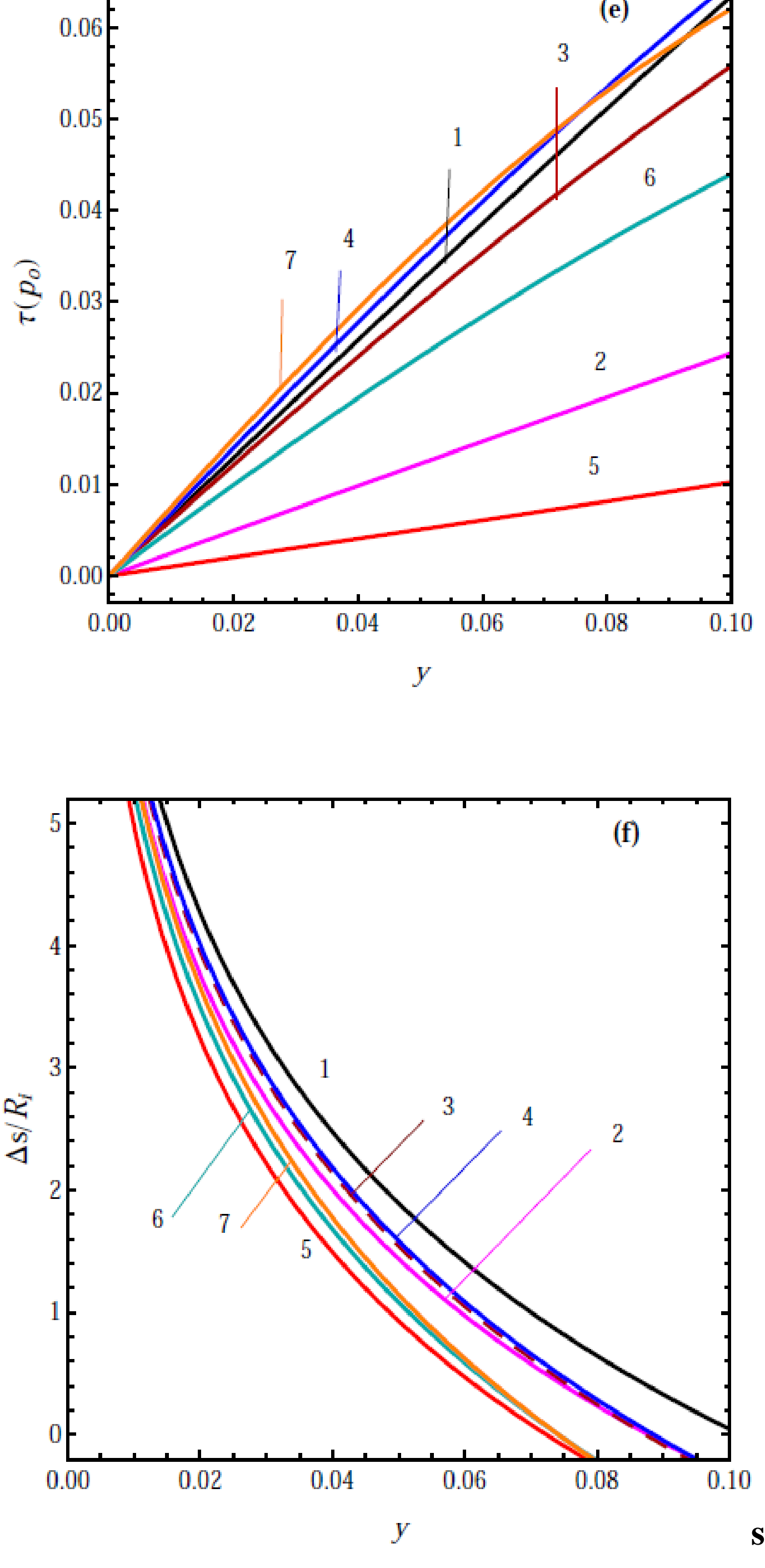

**Fig. 2** Variations of non-dimensional (a) velocity of fluid $u/a_o$, (b) pressure $p/p_o$, (c) density $\rho/\rho_o$, (d) speed of sound $a/a_o$, (e) adiabatic compressibility $\tau(p_o)$ and (f) change-in-entropy $\Delta s/R_i$ just behind the spherical shock front with the reciprocal square of Mach number $y$ for $\beta_{sp}=1$, $\gamma=7/5$, $x=0.975$ and various values of $k_p$ and $G$. **1**: $k_p=0$; **2**: $k_p=0.2, G=1$; **3**: $k_p=0.2, G=10$; **4**: $k_p=0.2, G=100$; **5**: $k_p=0.4, G=1$; **6**: $k_p=0.4, G=10$; **7**: $k_p=0.4, G=100$.

Figs. 2a, d, f). However, the behavior of the pressure, the density and the compressibility, in particular for $k_p = 0.4$, $G = 100$ differs greatly from the ideal gas (see Figs. 2b, c, e). The effects of an increase in the value of the parameter $G$ are: (i) to decrease the distance of piston from the shock front (see Table 1), (ii) to increase the strength of shock (see Fig. 1b–c), and (iii) to increase the velocity of fluid, the density, the adiabatic compressibility and the change-in-entropy, however, to decrease the pressure and the speed of sound (see Fig. 1). This behavior of the velocity of fluid, the density, the adiabatic compressibility and the change-in-entropy (see Fig. 1a, c, e, f), in particular for $k_p = 0.4$, $G = 1$ differs much more from the ideal gas. And also the variations of the pressure and the speed of sound (see Fig. 1b, d), especially for $k_p = 0.4, G = 100$ differ much more from the ideal gas. Obviously, the above effects are more impressive at higher value of the parameter $k_p$. The effects of an increase in the value of the parameter $k_p$ are as follows: (i) to increase the distance of piston from the shock front when $G = 1$. At higher values of the parameter $G$, the effect is small and of opposite nature (see Table 1), (ii) to decrease the velocity of fluid, the density, the compressibility when $G = 1$, and to increase them when $G = 100$, (iii) to increase the speed of sound, when $G = 1$, and to decrease when $G = 100$, and (iv) to decrease the pressure and the change-in-entropy.

**Table 1** Values of $x_p$, $J_0$, $\xi_0$, $\xi_1$, $\xi_2$, $\sigma_1$ and $\sigma_2$ for some typical values of $k_p$ and $G$.

| $k_p$ | $G$ | $x_p$ | $J_0$ | $\xi_0$ | $\xi_1$ | $\xi_2$ | $\sigma_1$ | $\sigma_2$ |
|---|---|---|---|---|---|---|---|---|
| 0 | | 0.916375 | 0.271372 | 1.292145 | 1.287308 | 14.44664 | 0.391209 | –16.58566 |
| 0.1 | 1 | 0.892291 | 0.237036 | 1.338631 | 1.979369 | 5.883546 | –0.444587 | –2.863075 |
| | 10 | 0.919720 | 0.276241 | 1.292266 | 1.430616 | 26.25784 | 0.407887 | –31.30408 |
| | 100 | 0.922396 | 0.281301 | 1.285913 | 1.291596 | 41.74705 | 0.606680 | –51.67292 |
| 0.2 | 1 | 0.865531 | 0.202362 | 1.396237 | 2.399740 | 2.801849 | –0.863860 | 3.308124 |
| | 10 | 0.922396 | 0.280595 | 1.292709 | 1.618171 | 35.44443 | 0.411433 | –42.42924 |
| | 100 | 0.929086 | 0.290809 | 1.282234 | 1.414504 | 36.82785 | 0.719783 | –44.94910 |
| 0.3 | 1 | 0.835426 | 0.168360 | 1.465942 | 2.892942 | –152.0125 | –1.343243 | 205.4876 |
| | 10 | 0.924403 | 0.283267 | 1.295831 | 1.947445 | 36.39492 | 0.292589 | –42.24241 |
| | 100 | 0.935776 | 0.301237 | 1.277442 | 1.532937 | 44.77832 | 0.915057 | –54.75224 |
| 0.4 | 1 | 0.801307 | 0.135028 | 1.554929 | 3.196870 | –3.899304 | –1.489377 | 17.27737 |
| | 10 | 0.925072 | 0.283783 | 1.301563 | 2.443695 | 34.73897 | 0.047813 | –37.49841 |
| | 100 | 0.942466 | 0.312653 | 1.271467 | 1.655133 | 79.33410 | 1.217995 | –98.71363 |

Obviously, the distance between the shock front and the inner expanding surface decreases with an increase in the value of the parameter $G$. As the value of the parameter $k_P$ increases the distance of piston from the shock front decreases at higher values of $G$, however, it increases when $G = 1$. The pressure and the density distributions, from the inner expanding surface to shock front, become steeper for higher values of the parameter $G$ (see Fig. 1b–c) which means the shock strengthens.

## Conclusions

This study presents the power series solutions for flow variables just behind the strong spherical shock waves of time dependent variable strength in a two-phase gas-particle medium. The following conclusions are drawn from the findings:

1. The velocity of fluid, speed of sound, adiabatic compressibility and change-in-entropy increase, however, the pressure and density decrease as we move towards the inner expanding surface.
2. The velocity of fluid, density and speed of sound increase, however, the distance between the shock front and the inner expanding surface, pressure, adiabatic compressibility and change-in-entropy decrease with increase in the parameter $k_p$.
3. The velocity of fluid, density, adiabatic compressibility and change-in-entropy increase, however, the distance between the shock front and the inner expanding surface, pressure and speed of sound decrease with increase in the parameter $G$.
4. The velocity of fluid, pressure, speed of sound and change-in-entropy decrease, however, the density and adiabatic compressibility increase with increase in $M^{-2}$.
5. The effects due to the dust-loading parameters, generally, do not change the trends of variations of the flow variables but they modify the numerical values of these flow quantities from their values for the ideal gas.
6. The trends of variations of the flow quantities are same in a two-phase gas-particle medium and ideal gas.

The present model is expected to facilitate to design some laboratory dusty plasma experiments which may observe the trends or variations of flow quantaties (behind the LASER induced shocks) that we predicted in this paper.